\documentclass[12pt,preprint2]{aastex}
\usepackage{emulateapj5,apjfonts,onecolfloat5}

\slugcomment{Accepted for publication in the Astrophysical Journal Letters} 

\shorttitle{New Constraints on G292.0+1.8 Containing PSR J1124-5916}
\shortauthors{Gonzalez and Safi-Harb}

\def\chan{$\it{Chandra}$}
\begin{document}
\twocolumn[

\title{New Constraints on the Energetics, 
Progenitor Mass, and Age of the Supernova Remnant G292.0+1.8 Containing PSR J1124-5916}
\author{Marjorie Gonzalez\altaffilmark{1} and Samar Safi-Harb\altaffilmark{1}}
\affil{$^{1}$Physics and Astronomy Department, University of Manitoba, Winnipeg, MB, R3T 2N2, Canada; 
umgonza4@cc.umanitoba.ca, safiharb@cc.umanitoba.ca.} 

\begin{abstract}
We present spatially resolved spectroscopy of the supernova remnant (SNR) G292.0+1.8 with
the  \chan\ X-ray observatory. 
This composite-type SNR contains the 135 ms pulsar, J1124-5916, recently discovered by Camilo and coworkers. 
We apply non-equilibrium ionization (NEI) models to fit the ejecta-dominated regions 
and to identify the blast wave of the supernova explosion. 
By comparing the derived abundances with those predicted from nucleosynthesis models, 
we estimate a progenitor mass of 30--40~M$_{\odot}$.  
We derive the intrinsic parameters of the supernova explosion such as its energy, 
the age of the SNR, the blast wave velocity, and the swept-up mass.
In the Sedov interpretation, our estimated SNR age of 2,600$^{+250}_{-200}$ years
 is close to the pulsar's characteristic age of 2,900 years. This confirms the pulsar/SNR association and 
relaxes the need for the pulsar to have a non-canonical value for the braking index, 
a large period at birth or a large transverse velocity. 
We discuss the properties of the pulsar wind nebula (PWN) in the light of the Kennel and Coroniti model 
and estimate the pulsar wind magnetization parameter.
We also report the first evidence for steepening of the power law spectral index with increasing radius
from the pulsar, a result that is expected from synchrotron losses and is reminiscent of
Crab-like PWNe. 
\end{abstract}


]

\section{Introduction}
\chan\ has been advancing
our understanding of core-collapse supernovae by revealing their collapsed cores
and their outflows, the distribution of supernova ejecta, and their interaction
with the interstellar medium (ISM). 
Increasing the sample of SNRs with a hybrid morphology
(i.e. Composite-type SNRs)
helps address questions related to the birth properties and evolution of
neutron stars and SNRs in the ISM.
\par
G292.0+1.8 (MSH~11-54) offers a unique laboratory to address these
questions because of its hybrid morphology, its
youth, and its size.
This SNR is one of three Oxygen-rich remnants in our Galaxy. It was first discovered in 
the radio band (Milne 1969; Shaver \& Goss 1970), 
where it exhibits a centrally-filled morphology 
indicative of the presence of a PWN.
$\it{Einstein}$ and $\it{EXOSAT}$ X-ray observations classified the remnant as a type-II explosion of a 
massive star (Hughes \& Singh 1994).
\chan\  observations led to the discovery of the long-sought PWN 
(Hughes et al. 2001; Safi-Harb \& Gonzalez 2001) surrounding a candidate pulsar. 
An imaging study of the SNR with \chan\
(Safi-Harb \& Gonzalez 2001; Park et al. 2002) 
traced the distribution of the ejecta.
The remnant has a morphology similar
to Cas~A on small scales (Hwang, Holt, \& Petre 2001),
however, no Fe-K  line-emission was detected.

The discovery of the PWN in G292.0+1.8
prompted a deep search for radio pulsations and led to the discovery of a 135 ms pulsar 
coincident with the \chan\ X-ray point source (Camilo et al. 2002a). 
The pulsar has a derived spin-down luminosity  $\dot{E}$ of 1.2$\times$10$^{37}$ erg s$^{-1}$, 
a surface magnetic dipole field strength of 1$\times$10$^{13}$ G, 
and a characteristic age of 2,900 yrs.
To date, the only estimate of the age of this SNR
is from optical studies of the high-velocity material
coincident with the PWN (Murdin \& Clark 1979).
The age of 1,600 years is a factor of 1.8 less than the characteristic
age of the pulsar.
\par
In this Letter, we present the first detailed spatially resolved spectroscopy of
the \chan\ observation  in order to infer 
the intrinsic parameters of the supernova explosion
and its associated pulsar and PWN.
In \S2, we briefly summarize the observation
and present our sub-arcminute spatially resolved spectroscopy
of the SNR and the embedded PWN. 
In \S3, we discuss our results and infer the energetics of the supernova explosion, the mass of the progenitor,
the SNR's age, the pulsar's transverse velocity and the pulsar wind parameter.
We also report the first evidence
for spectral softening of the power law index of the PWN with
increasing distance from the pulsar.

\section{Sub-arcminute spatially resolved spectroscopy}
G292.0+1.8 was observed with the S3-chip of the
Advanced CCD Imaging Spectrometer (ACIS) on board \chan\ on March 11, 2000, at a 
CCD temperature of -120$^o$C. 
We corrected for Charge Transfer Inefficiency (CTI) by applying the $\it{cticorrectit}$
tool to the event~1 raw data (Townsley et al. 2000). 
A new event 2 file was then obtained by screening the data 
using standard ${\it Ciao~2.2}$ routines. 
 The resulting exposure time was 38 ksec. 

\subsection{SNR}
To perform spatially resolved spectroscopy of the SNR, we selected the 23 bright regions 
shown in Fig.~1 (left). In this image, 
the broadband emission from the SNR was divided into soft (0.3--1.15 keV, red), 
medium (1.15--2.15 keV, green) and hard (2.15--10.0 keV, blue) contributions, 
each smoothed with a gaussian with $\sigma$=1\farcs5 . The PWN is clearly
visible in blue near the center of the SNR.
We fitted the spectra of the remnant in the 0.5--8.0 keV range using
the non-equilibrium ionization (NEI) model, 
$\it{vpshock}$ (Borkowski, Lyerly, \& Reynolds 2001) in version 11.0.1 of the XSPEC package
(Arnaud 1996).
This model is appropriate for modeling 
plane-parallel shocks in young SNRs whose plasma has not reached ionization equilibrium. It
is characterized by the shock temperature, $kT$, and the ionization timescale, $n_{e}t$, of the plasma. 
Here, $n_{e}$ is the post-shock electron density and $t$ is the time since the passage of the shock. 
$t$ can then be used as a lower estimate on the SNR age. Since the remnant is ejecta-dominated,
we allowed the abundances to vary.
\par
One-component NEI model fits
yielded an average temperature for the selected regions of 0.9$\pm$0.3 keV, 
with minimum and maximum values of 0.6 and 1.7 keV, respectively. 
The ionization timescale varied across the SNR from $\sim$10$^{11}$ to 10$^{13}$ cm$^{-3}$~s. 
These values indicate that the emission from the ejecta contains
a wide range of temperatures and ionization states.
The highest ionization state values are associated with regions 1, 3, and 9 (see Fig. 1, left),
indicating that the plasma has reached ionization equilibrium in the southeast.
Relative abundances were well constrained in these models and were used to derive the abundance ratios 
discussed in \S3.2.

\par
We subsequently attempted 
two-component model fits.
This was motivated by the expectation of 
a high-temperature and a low-temperature
  plasma associated with the supernova blast wave and reverse shock, respectively
(see e.g. Safi-Harb et al. 2000).
While the overall fits were slightly improved,
the abundances could not be well constrained due to the large number of free parameters.
The derived average temperatures for the hot and cool components were 
1.05$\pm$0.34 keV and 0.37$\pm$0.18 keV, respectively.
These are lower than 1.64$^{+0.3}_{-0.2}$ keV, as reported by Hughes \& Singh (1994) 
from their one-component NEI model fit to the $\it{Einstein}$ and $\it{EXOSAT}$ data. 
The discrepancy is probably due to the contamination by the previously unresolved PWN. 

\subsection{PWN}
\chan\ observations of PWNe have shown that 1)~PWNe harbor torus-like and
jet-like structures indicating equatorial and polar outflows,
and 2) the power law photon index steepens away from the
pulsar, as expected from synchrotron and expansion losses (see e.g. Safi-Harb 
2002 for a review).
By examining this PWN, we find evidence of an arcsecond-scale east-west elongation and an extension
to the south (see Fig~1, right), hinting at structures associated with the
deposition of the pulsar's wind energy into its surroundings.
To search for spectral variations, we divided the inner $45''$ of the PWN into four concentric rings 
to obtain enough counts to extract a spectrum.
 As reported by Hughes et al. (2001), the spectrum of the
PWN is heavily contaminated by thermal emission from the SNR.
In order to minimize this contamination,
we used a nearby region as background and restricted the fit to the 1.5--7.0 keV range,
thus reducing the contamination by Oxygen, Neon, and Magnesium ejecta.
Using a  power-law model, we find that the photon index 
($\Gamma$) steepens away from the pulsar (see Table~1). This result is expected
from synchrotron losses and is reminiscent of Crab-like nebulae (e.g. Safi-Harb et al. 2001).

\section{Discussion}

\subsection{Distance}
From our one- and two-component fits, we derive an average value of 
$N_{H}$=(0.5 $\pm$ 0.1)$\times$10$^{22}$ cm$^{-2}$ for the remnant. 
We can use this value to verify previous estimates for the distance to G292.0+1.8. 
The extinction per unit distance in the direction of the remnant can be estimated from
the contour diagrams given by Lucke (1978): $E_{B-V}/D$$\sim$0.2 mag~ kpc$^{-1}$. 
Using  the relation $N_{H}/A_{V}$=1.79$\times$10$^{21}$~cm$^{-2}$~mag$^{-1}$, 
which translates into $N_{H}/E_{B-V}$=5.55$\times$10$^{21}$~cm$^{-2}$~mag$^{-1}$ 
(Predehl \& Schmitt 1995), we derive a distance of 3.6--5.5 kpc. 
A distance of 4.8 kpc, as used in the previous studies, is then 
a good approximation and is adopted in the following discussion. 

\subsection{Supernova Parameters}
We subsequently use our NEI model fits to derive the supernova parameters.
In the standard picture of X-ray emission from young SNRs, the soft
component arises from shocked ejecta and the hard component is usually
attributed to the blast wave shocking the surrounding medium.
Our spatially resolved spectroscopy indicates that most of the
regions in the SNR (except for the PWN, the outermost regions and the eastern 
`belt'-like filament near the SNR center)
are well fitted with a one-component  NEI ($\it{vpshock}$)
model with  high metal abundances (with respect to solar),
indicating that these knots are the shocked
ejecta associated with the explosion of a massive star.
Nucleosynthesis yield as a function of the progenitor's
mass has been numerically computed by Woosley and Weaver (1995) 
and several others.
In order to derive the progenitor's mass, 
we use the abundance ratios from our NEI model fits in the 0.5--8.0 keV band
to the ejecta-dominated regions.
In Fig.~2, we plot the average values for these ratios along with their rms scatter. 
Over-plotted are the predicted ratios from nucleosynthesis models
for progenitors of 25, 30, 35 and 40~M$_{\odot}$.
We then conclude that the measured abundance ratios are consistent with a progenitor's
mass of 30--40~M$_{\odot}$.

Regions 20 to 23 selected in our analysis are of special interest (see Fig.~1, left). 
Region 20 is part of a set of sharp circumferential filaments surrounding the bulk 
of the ejecta-dominated X-ray emission. 
Regions 21, 22 and 23 are part of an outer faint ridge of emission delimiting the SNR. 
We believe that these latter regions indicate the location of the blast wave shocking the
surrounding medium for the following reasons:
1)~as shown in Table~2, the abundances derived for region 21 using one-component models 
are consistent with ISM abundances (sub-solar) and are the lowest ones throughout the remnant,
while those derived for region 20 are contaminated by ejecta;
2)~the outermost emission is absent in the equivalent width maps which reveal the ejecta (Park et al. 2002); 
and 3)~even two-component model fits to regions 21--23 require sub-solar abundances, which is not the case 
for region 20, where both components require 
above-solar abundances.

In order to estimate the SNR age, we will adopt our more reliable one-component fits 
to regions 21--23 in the 0.5--8.0 keV band. The corresponding SNR size would be r$\sim$$4.2'$=5.9 pc 
(at 4.8 kpc). In the following example, we will use the parameters derived for region 21 shown in 
Table 2. The observed emission measure (EM) of 0.09$\pm$0.02 
(after scaling it to the SNR volume) corresponds to:
$\int$$n_{e}$$n_{H}$dV=10$^{14}$(4$\pi$D$^{2}$)(EM) = (2.5$\pm$0.6)$\times$10$^{58}$ cm$^{-3}$.
If the emission volume V=$f$$V_{tot}$, where $f$ is the filling factor, we find
an upstream density $n_{0}$ = (0.23$\pm$0.03) $f^{-1/2}$~cm$^{-3}$. 
From our fit parameter $n_{0}t$, obtained from the $\it{vpshock}$ model, we find
a shock age of $t$=(2,500$^{+1100}_{-700}$)$f^{1/2}$ years.
Similarly, we derive the parameters for regions 22 and 23 and summarize them in Table 3. 
For a reasonable filling factor of 0.25, 
the shock age varies from 1,250--4,450 years.
The discrepancy between these ages suggests that the northeastern regions are shocked
at earlier times that the western regions, probably due to an inhomogeneous medium.
This complication makes the determination of the SNR parameters in this manner uncertain.

\par

In order to estimate the global properties
of the SNR independently of the small-scale density variations, we
attempted to fit the spectrum of the whole remnant.
We found this to be extremely difficult using the entire 0.5--8.0 keV band 
and various combination models with up to three components.
This difficulty is due to the large number of NEI components needed to account
for the variations in temperature and ionization time-scale of the ejecta,
as indicated in our spatially resolved spectroscopy described above.
However, restricting the fit to the 2.0--8.0 keV range (to remove the contamination
from O, Ne, Mg and Si ejecta) and excluding the pulsar and its PWN,
we found that a two-component model ($vpshock+sedov$)
provides an acceptable fit for the overall spectrum with $\chi$$^{2}_{\nu}$(dof)=1.17(197);
$\it{sedov}$ is an NEI model that
accounts for the hot thermal emission from the SNR interior
and is based on the Sedov dynamics (Borkowski et al. 2001).
The $\it{vpshock}$ model was used to characterize the emission from 
the Sulfur ejecta, while the $sedov$ model
was used to characterize the blast wave component 
(abundances of 2/3 solar with equal electron and ion temperatures).
In Table~4, we summarize the SNR parameters estimated from 
using equations (4a-4f) in Hamilton et al. (1983).
The derived tabulated value of $n_0$ is consistent within the various uncertainties
with that derived from the observed emission measure.
The SNR age, $t$(Sedov)=2,400--2,850 yrs, is comparable 
to the average value derived from regions 21--23 above. This age represents a more 
reliable estimate for the age of the remnant since the contamination from ejecta
has been minimized and the effect of an inhomogeneous medium
has been averaged. In addition, this age is in close agreement with the 
pulsar's characteristic age of 2,900 years.
This agreement strengthens the pulsar/SNR association
and relaxes the need for invoking a very slow initial spin 
period or a non-canonical value for the braking
index of the pulsar (Camilo et al. 2002a).

\subsection{Pulsar Wind Nebula Parameters}
As noted by Hughes et al. (2001),
the pulsar and its PWN are not located at the geometrical center of the remnant. 
The offset was attributed to either a high transverse velocity for the pulsar 
or slower expansion of the SNR toward the south-east. 
From the offset ($\sim$$1'$) they calculate a transverse velocity for 
the pulsar of $\sim$770($d_{4.8}/t_{1600}$)~km~s$^{-1}$, 
where $d_{4.8}$ is the distance in units of 4.8 kpc and $t_{1600}$ 
is the age of the remnant in units of 1,600 yrs. 
Using our new estimate of 2,600$^{+250}_{-200}$ yrs for the age we obtain a 
transverse velocity for the pulsar of 470$\pm$40~$d_{4.8}$~km~s$^{-1}$. 
This value is in excellent agreement with recent estimates of the mean birth velocity of pulsars 
(450$\pm$90 km s$^{-1}$, Lyne \& Lorimer 1994). 
The nearly east--west extension
and jet-like structure extending to the south (see Fig. 1, right)
could be indicative of a toroidal and
collimated outflows reminiscent of Crab-like PWNe.

In the Kennel \& Coroniti (1984) model, 
the synchrotron nebula is produced by the confinement of a 
relativistic wind  injected by the pulsar into its surroundings. 
The radius of the standing shock, $R_{s}$, is estimated by equating the pressure of the pulsar's wind, 
$\dot{E}$/(4$\pi$c$R^{2}_{s}$), with the 
pressure in the nebula, $P_{n}$=$B_n^2$/$4\pi$.
$B_{n}$, the nebular magnetic field,
can be approximated from its equipartition value, $B_{n, eq}$.
Using the average luminosity and spectral index for the PWN,
we estimate a value of $B_{n, eq}\sim$1$\times$10$^{-4}$~G. 
With the measured $\dot{E}$=1.2$\times$10$^{37}$ ergs s$^{-1}$ (Camilo et al. 2002a), 
we obtain a shock radius of $R_{s}$$\sim$$3''$ at 4.8 kpc.
Alternatively, we can estimate the shock radius using the
thermal pressure in the SNR interior as a confining pressure.
This average pressure can be estimated as 
2$n_e$$kT$ $\sim$ 6.9$\times$10$^{-9}$ erg~cm$^{-3}$;
which yields a shock radius of $1''$ .
The inner circle shown in Fig.~1 (right) shows the estimated upper value of $R_{s}$. 
An interesting parameter than can also be derived from \chan\ studies of PWNe is the
pulsar wind magnetization parameter, $\sigma$,
defined as the ratio of the Poynting flux to the particles flux, 
which can be estimated as $\sigma\sim(R_{s}/R_{n})^{2}$.
Using a shock radius $R_s$$\sim$$1''$--$3''$ and 
a nebular radius $R_{n}$$\sim$$30''$--$45''$,
we find that $\sigma$$\sim$5$\times$10$^{-4}$--0.01, indicating
a particle dominated wind.

The properties of the PWN can thus be derived from the Kennel \& Coroniti model.
The value derived for the pulsar wind magnetization parameter in this manner is 
comparable to that derived for other PWNe of similar properties (Safi-Harb 2002).
It is worth noting that the properties of the pulsar and the PWN in G292.0+1.8 are
strikingly similar to those of the pulsar discovered in the Crab's cousin,
G54.1+0.2 (Camilo et al. 2002b, Lu et al. 2002).
The lack of ejecta and a shell in the latter SNR (and in a handful other plerions
of comparable age) indicates that the classification of plerionic-type SNRs 
is to a large extent linked to the nature of the progenitor star, the
explosion energy, and the environment. 

\acknowledgments
This research made use of  NASA's
Astrophysics Data System (ADS), and of archival data obtained 
from the \chan\ X-ray Center.
We thank G.G. Pavlov for his help at the early stage
of this work, and for useful comments on
the manuscript. The authors acknowledge support by the Natural Sciences and Engineering
Research Council of Canada (NSERC).
M.E.G. is a PGS~A fellow, and SS-H is a University Faculty Award (UFA) fellow.


\onecolumn

\begin{deluxetable}{lcc}
\tablewidth{0pt} 
\tablecaption{Steepening of the power law photon index of the PWN in G292.0+1.8\tablenotemark{a}}
\tablehead{ \colhead{Radius} &\colhead{$\Gamma$ ($\pm$ 90\%)} & \colhead{$\chi^{2}$(dof)}}
\startdata
$<2''$ (pulsar)	&1.74$\pm$0.10	&142 (140)\\
$2''$ -- $7''$	&2.0$\pm$0.10	&122 (94)\\
$7''$ -- $15''$	&2.06$\pm$0.11	&129 (125)\\
$15''$ -- $30''$	&2.37$\pm$0.10	&202 (219)\\
$30''$ -- $45''$	&3.00$\pm$0.20	&204 (212)\\
\enddata
\tablenotetext{a}{$N_H$ fixed to the best fit value for the pulsar of 0.37$\times$10$^{22}$~cm$^{-2}$} 
\end{deluxetable}


\begin{deluxetable}{lcc}
\singlespace
\footnotesize
\tablewidth{0pt} 
\tablecaption{ One- and two-component NEI fits to regions 20 and 21}
\tablehead{ \colhead{} &\colhead{Region 20}	& \colhead{Region 21}}
\startdata
$N_{H}$ (10$^{21}$cm$^{-2}$) &4.7 $\pm$ 0.2  &4.9 $\pm$ 0.3 \\
$kT$ (keV)	&0.66$^{+0.01}_{-0.03}$		&0.87$\pm$0.07 \\
$n_{e}t$ (cm$^{-3}$~s)   &(7.5$^{+2.4}_{-1.8}$)$\times$10$^{11}$   & (8.7 $\pm$ 2.4)$\times$10$^{10}$  \\
O ($M_{\odot}$)	&1.1 $\pm$ 0.1	&0.38$^{+0.04}_{-0.06}$ \\
Ne ($M_{\odot}$)	&0.90 $\pm$ 0.06	&0.53$^{+0.05}_{-0.06}$ \\
Mg ($M_{\odot}$)	&0.63$^{+0.05}_{-0.06}$	&0.23 $\pm$ 0.06 \\
Si ($M_{\odot}$)	&0.47 $\pm$ 0.06	&0.27$^{+0.09}_{-0.08}$ \\
S ($M_{\odot}$)	&1.2 $\pm$ 0.3 	&0.2$^{+0.4}_{-0.2}$ \\
Fe ($M_{\odot}$)	&0.24 $\pm$ 0.02	&0.16 $\pm$ 0.03 \\
norm 	&(3.3 $\pm$ 0.2)$\times$10$^{-3}$   &(1.4 $\pm$ 0.2)$\times$10$^{-3}$ \\
$\chi$$^{2}_{\nu}$ (dof)	&1.9 (119)	&1.2 (98)\\
\cline{1-3}
$N_{H}$ (10$^{21}$ cm$^{-2}$)	&6.1 $\pm$ 0.03	&6.1$^{+0.02}_{-0.01}$ \\
$kT_{hot}$ (keV)	&0.94$^{+0.05}_{-0.02}$	& 0.95$^{+0.20}_{-0.15}$ \\
$n_{e}t_{hot}$ (cm$^{-3}$~s)	&(3 $\pm$ 0.5)$\times$10$^{11}$	& (6$^{+1.4}_{-1.8}$)$\times$10$^{9}$ \\
norm$_{hot}$	&(1.05 $\pm$ 0.05)$\times$10$^{-3}$   &(9.1 $\pm$ 1.4)$\times$10$^{-4}$ \\
$kT_{cool}$	(keV) &0.27 $\pm$ 0.02	 &0.84 $\pm$ 0.06 \\
$n_{e}t_{cool}$ (cm$^{-3}$~s)	&(2.7 $^{+2.3}_{-1.1}$)$\times$10$^{11}$ & (1.4 $\pm$ 0.3) $\times$10$^{11}$ \\
norm$_{cool}$	&(5.1 $\pm$ 0.4)$\times$10$^{-3}$   &(2.8$^{+0.3}_{-0.2}$)$\times$10$^{-4}$ \\
$\chi$$^{2}_{\nu}$ (dof)	&1.6 (116)	&1.0 (95) \\
\enddata
\end{deluxetable}


\begin{deluxetable}{lccc}
\singlespace
\footnotesize
\tablewidth{0pt} 
\tablecaption{Derived SNR parameters from Regions 21, 22 and 23}
\tablehead{ \colhead{} &\colhead{Region 21}	& \colhead{Region 22}& \colhead{Region 23} }
\startdata
Ionization timescale, $n_{e}t$ (cm$^{-3}$~s) &(8.7$\pm$2.4)$\times$10$^{10}$ & (2.1$\pm$0.6)$\times$10$^{11}$ & (3.1$^{+1.2}_{-0.6}$)$\times$10$^{11}$ \\
$\int$$n_{e}$$n_{H}$dV (cm$^{-3}$)	&	(2.5$\pm$0.6)$\times$10$^{58}$ & (3.9$\pm$0.9)$\times$10$^{58}$ & (2.5$\pm$0.8)$\times$10$^{58}$ \\
Ambient density, $n_{0}$ ($f^{-1/2}$~cm$^{-3}$) & 0.23$\pm$0.03 & 0.29$\pm$0.4 & 0.23$\pm$0.04 \\
Swept-up mass, $M_{s}$ ($f^{-1/2}$~M$_{\odot}$) & 6.7$\pm$0.8 & 8.3$\pm$0.9 & 6.7$\pm$1.1 \\
Lower age limit, $t$ ($f^{1/2}$~years)	& 2,500$^{+1,100}_{-7,00}$ & 4,900$^{+2,100}_{-1,500}$ & 8,900$^{+5,700}_{-2,800}$ \\
\enddata
\end{deluxetable}


\begin{deluxetable}{lc}
\tablewidth{0pt} 
\tablecaption{Explosion parameters from $sedov$ fit to hard component of the overall spectrum}
\tablehead{ \colhead{Parameter} &\colhead{Value ($\pm$ 1$\sigma$)}}
\startdata
Emission Measure, $EM$ & 0.166$^{+0.006}_{-0.008}$ \\
Shock temperature, $kT_{s}$ (keV)	&0.95 $\pm$ 0.15 \\
Ionization timescale, $n_{e}t$ (cm$^{-3}$~s) &(2.0$^{+0.4}_{-0.3}$)$\times$10$^{11}$\\
Shock velocity, $v_{s}$ (km~s$^{-1}$)  & 880 $\pm$ 70\\
Age, $t$ (yrs)	& 2,600$^{+250}_{-200}$	\\
Ambient density, $n_{0}$ (cm$^{-3}$)  &0.51$^{+0.15}_{-0.11}$\\
Explosion energy, $E_{0}$ (10$^{51}$ ergs)   & 0.18$^{+0.08}_{-0.06}$\\
Swept-up mass, $M_{s}$ (M$_{\odot}$)	& 15.6$^{+4.5}_{-3.5}$ \\
\enddata
\end{deluxetable}


\begin{figure}
\begin{center}
\centerline {\includegraphics[scale=.94]{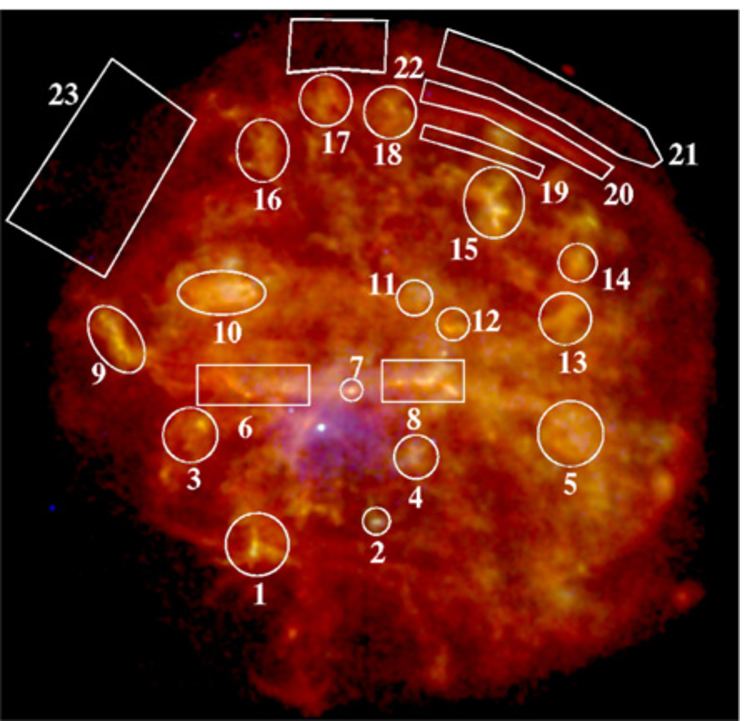}  \includegraphics[scale=1]{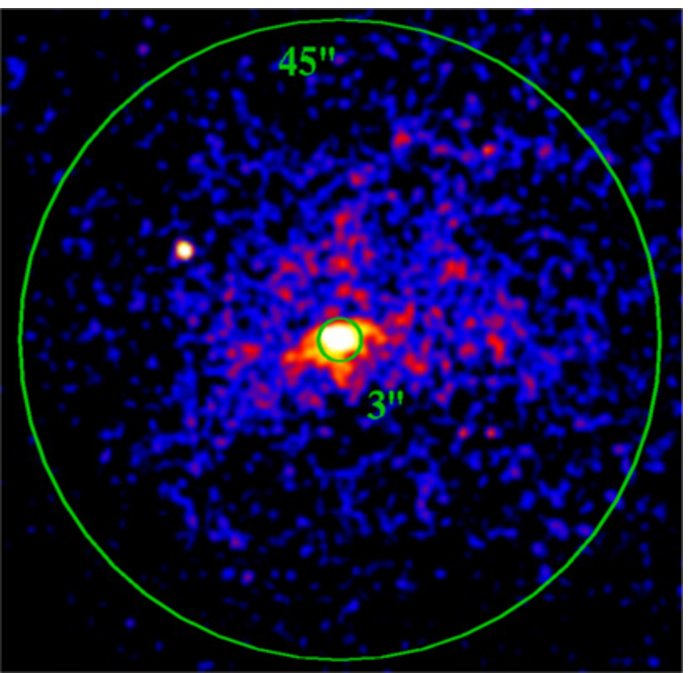}}
\caption{\label{1} Left: combined broadband image of G292.0+1.8;
individual images were used from the soft (0.3--1.15~keV, red), medium
(1.15--2.15 keV, green) and hard (2.15--10.0 keV, blue) bands each
smoothed with a gaussian with $\sigma$=$1\farcs5$. Superimposed are the regions
used for our spatially resolved spectroscopy. Right: 2.6--10.0 keV false color image
of the PWN surrounding PSR J1124-5916. The image was smoothed
using a gaussian with $\sigma$ = $0\farcs25$. The inner circle is at a radius of $3''$ 
($R_s$) and the outer one is at $45''$ ($R_n$).}
\end{center}
\end{figure}

\begin{figure}
\begin{center}
\centerline {\includegraphics[scale=1]{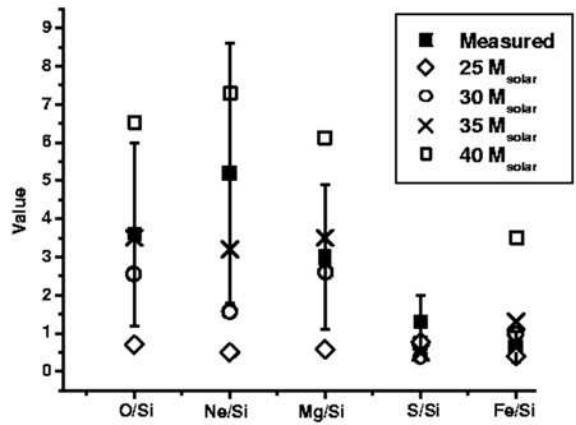}}
\caption{\label{2} Average abundance ratios (filled squares) and their rms
scatter measured throughout the remnant. Predicted ratios from
nucleosynthesis models are also shown.}
\end{center}
\end{figure}

\end{document}